\documentclass{article}
\usepackage{times}
\usepackage{amsfonts}
\usepackage{graphicx}
\begin{document}
\noindent
{\Large  SPHERICAL SPACE IN THE NEWTONIAN LIMIT:\\THE COSMOLOGICAL CONSTANT}
\noindent

\vskip.5cm
\noindent
{\bf  J.M. Isidro}$^{a}$, {\bf P. Fern\'andez de C\'ordoba}$^{b}$ and {\bf J.C. Castro-Palacio}$^{c}$\\
Instituto Universitario de Matem\'atica Pura y Aplicada,\\ Universitat Polit\`ecnica de Val\`encia, Valencia 46022, Spain\\
${}^{a}${\tt joissan@mat.upv.es}, ${}^{b}${\tt pfernandez@mat.upv.es},\\
${}^{c}${\tt juancas@upvnet.upv.es}\\

\vskip.5cm
\noindent
\today
\vskip.5cm
\noindent
{\bf Abstract}  We compute the cosmological constant of a spherical space in the limit of weak gravity.  To this end we use a duality developed by the present authors in a previous work. This duality allows one to treat the Newtonian cosmological fluid as the probability fluid of a single particle in nonrelativistic quantum mechanics. We apply this duality to the case when the spacetime manifold on which this quantum mechanics is defined is given by $\mathbb{R}\times\mathbb{S}^3$. Here  $\mathbb{R}$ stands for the time axis and  $\mathbb{S}^3$ is a 3--dimensional sphere endowed with the standard round metric. A quantum operator $\Lambda$ satisfying all the requirements of a cosmological constant is identified, and the matrix representing $\Lambda$ within the Hilbert space $L^2\left(\mathbb{S}^3\right)$ of quantum states is obtained. Numerical values for the expectation value of the operator $\Lambda$ in certain quantum states are obtained, that are in good agreement with the experimentally measured cosmological constant.

\tableofcontents

\section{Introduction}\label{einfuehrung}

Newtonian cosmology succeeds in capturing some essential physics of the Universe (for a very incomplete sample see, {\it e.g.}\/, refs. \cite{BARROW1, BENISTY, BONDI, CABRERA, NOSALTRES}). It also has the advantage that its mathematics is notably simpler than that of its general--relativistic parent theory.

At the same time, the cosmological constant problem remains one of the deepest mysteries in theoretical physics (for reviews see, {\it e.g.}\/, refs. \cite{CARROLL, MARTIN, PADDY0, PADDY1, PEEBLES, WEINBERG2}). One hopes that an eventual theory of quantum gravity will shed light on this foundational question. However, for as long as a theory of quantum gravity remains elusive, one has to make do with less ambitious but well established theoretical frameworks. Both general--relativistic cosmology and its weak--gravity limit, Newtonian cosmology, count as such.

In this article we present a toy model of Newtonian Universe in positively curved space, in which a certain repulsive force plays a role analogous to that played by the cosmological constant $\Lambda$ in general relativity. This follows previous work by the same authors \cite{NOSALTRES}, where we have computed the cosmological constant of flat, Newtonian space. In the present paper we extend this previous analysis to the case of positive (but constant) scalar curvature, still within the limit of weak gravity. 

The 4--dimensional spacetime manifold considered here will be $\mathbb{R}\times\mathbb{S}^3$. The real line will account for the time variable, while the sphere $\mathbb{S}^3$ will be endowed with the standard round metric. Then $\mathbb{R}\times\mathbb{S}^3$ becomes the 4--manifold considered by Einstein in his proposal for a static Universe (as reported, {\it e.g.}\/, in ref. \cite{TOLMAN}). Of course, this cosmological model has long been discarded for running contrary to well--established experimental facts:  it fails to exhibit redshifts because it is static instead of expanding \cite{HUBBLE, PERLMUTTER, RIESS}, and it has positive curvature instead of being flat. (The value of the spatial curvature of our Universe remains the subject of ongoing debate \cite{VAGNOZZI}, though). Still, this model is worth revisiting from the viewpoint of current trends such as, {\it e.g.}\/, the modern thermodynamic approach to gravity and spacetime \cite{CABRERA, NOSALTRESTRIS, PADDY1, VERLINDE}, which is the standpoint adopted here. 

We refer the reader to our paper \cite{NOSALTRES} for a thorough explanation of our use of nonrelativistic quantum mechanics in order to model the Newtonian cosmological fluid. Since this feature is, to the best of our knowledge, new in the literature, let us reiterate: {\it we will mimic the Newtonian cosmological fluid as the quantum mechanics of a single nonrelativistic particle within a given spacetime manifold}\/. In particular, the density of matter across the Universe is (proportional to) the probability density $\vert\psi\vert^2$ of the quantum mechanics described by the wavefunction $\psi$. Thus $\int_V{\rm d}^3x\vert\psi\vert^2$ is (proportional to) the total mass within the volume $V$. The particle described by $\psi$ will be assumed free, as already done in ref. \cite{NOSALTRES}. 

A welcome geometrical feature of spherical geometry is that one can identify a {\it radial}\/ coordinate orthogonal to the remaining angular coordinates; the latter parametrise a 2--dimensional sphere $\mathbb{S}^2$ in the case of 
$\mathbb{S}^3$. In our finite--dimensional  quantum--mechanical model an operator $\Lambda$ can be identified, such that it plays the role of the cosmological constant. Specifically, the sought--for operator $\Lambda$ is the inverse of the square of the radial distance $\rho$, the latter measured with respect to the given metric on the spatial manifold $\mathbb{S}^3$:
\begin{equation}
\Lambda=\frac{C}{\rho^2}.
\label{landaerre}
\end{equation}
Above, $C$ is a numerical constant, the precise value of which will be specified presently. This constant is dimensionless because $\rho$ carries the dimensions of length. 

Let us motivate our choice (\ref{landaerre}) in detail.\footnote{We thank a referee for demanding the clarifications presented here.}\\
{\it i)} To begin with, it is dimensionally correct: it carries the dimension of inverse length squared, as befits the cosmological constant (in nonnatural units).\\ 
{\it ii)} We have in ref. \cite{NOSALTRES} made use of the flat--space analogue of the operator (\ref{landaerre}), to successfully apply the same programme developed here: a determination of the cosmological constant of a flat, Newtonian Universe given by $\mathbb{R}\times \mathbb{R}^3$. As argued there, one can interpret the cosmological constant as a repulsive centrifugal force described by an inverse--square potential function given by Eq. (\ref{landaerre}).\\ 
{\it iii)} Eq. (\ref{landaerre}) expresses a quantum operator. As such, the result of a measurement of $\Lambda$ can only be one of the eigenvalues of this operator. Since {\it a priori}\/ we do  not know the quantum state that our (hypothetical)  Universe $\mathbb{R}\times\mathbb{S}^3$ finds itself in, we will proceed in reverse: from a knowledge of the experimentally measured value of $\Lambda$ we will identify the quantum eigenstate whose eigenvalue fits best. \\
{\it iv)} Unfortunately we do not believe the different modes (quantum eigenstates) associated with the quantum operator (\ref{landaerre}) can be detected; there are two reasons for this. First, our actual Universe appears to be flat rather than positively curved (even if this point remains unsettled; see ref. \cite{VAGNOZZI}). Second, for an eventual detection of such modes one would require a corresponding {\it spectroscopy}\/, as in atomic and nuclear theory. As it is, we have just one experimentally measured value of the cosmological constant, with no indications of transitions between different states corresponding to different values of $\Lambda$. This forces us to apply the {\it reverse procedure}\/ mentioned above, that is, to look for the best fit to a given value of $\Lambda$. All this notwithstanding, the spherical modes used here have found considerable applications in other cosmological problems; for a sample see, {\it e.g.}\/, ref. \cite{LINDBLOM}.\\
{\it v)} The use of a quantum operator to measure a classical quantity may appear surprising. However, we will verify {\it a posteriori}\/ that the above--mentioned eigenvalue, or even the quantum expectation value of $\Lambda$ in a state other than an eigenstate, will have semiclassical properties. This will be the case, {\it e.g.}\/, whenever the quantum numbers of the states involved are large. Indeed we will verify, at the end of our calculations, that all states involved will carry large quantum numbers, so they all enjoy semiclassical properties.  Based on the duality mentioned two paragraphs above, we simply find it convenient to represent Newtonian cosmological quantities by means of nonrelativistic quantum mechanics. From this point of view, the use of the quantum--operator formalism in order to compute classical quantities is totally justified.\\
{\it vi)} The cosmological constant $\Lambda$ is being represented by a coordinate--dependent operator. We are working in the Newtonian limit, where diffeomorphism invariance is not an essential requirement of the theory. Moreover, most quantum--mechanical operators are actually coordinate dependent.\\ 
{\it vii)} Our choice (\ref{landaerre}) involves one specific coordinate, namely the radial coordinate, that is naturally present in the sphere $\mathbb{S}^3$ (see Eqs. (\ref{uno})--(\ref{catorce})). Since the latter is a homogeneous manifold, the radial coordinate that is being used to define (\ref{landaerre}) can always defined around any point of the sphere, with no changes whatsoever to the spectral properties of the operator $\Lambda$. Here we are making use of the high symmetry of the spatial manifold $\mathbb{S}^3$ to counter the false impression that results based on our choice (\ref{landaerre}) might be flawed  because of lack of covariance.\\
{\it viii)} In ref. \cite{MILKYWAY} it has been demonstrated that the $\Lambda$ contribution in a spherically symmetric solution changes the mass of M31and the Milky Way. Since the mass of the local group is dominated by dark matter, which cannot be observed directly, an exact determination of the role played by $\Lambda$ becomes crucial. We hope that the techniques presented here, possibly with some modifications, may eventually be used in a determination of the mass of the local group. The modifications required surely include (but are not limited to) taking the large--radius limit (the flat--space limit) of the sphere $\mathbb{S}^3$. 

Having now justified our choice (\ref{landaerre}), we turn to a description of our programme. We will first solve the time--independent  Schroedinger equation for a free particle of mass $M$ on the spatial manifold $\mathbb{S}^3$. The Laplacian operator $\nabla^2$ will be the one dictated by the Riemannian metric on $\mathbb{S}^3$; the free Schroedinger equation is in fact equivalent to the eigenvalue equation for the Laplacian operator on $\mathbb{S}^3$. We will therefore use the terms ``Laplacian eigenfunctions" and ``Schroedinger eigenfunctions" interchangeably. However, in order to conform to standard usage, the free Schroedinger equation will be written $\nabla^2\psi$ $+\kappa^2\psi=0$ with $\kappa^2=2ME/\hbar^2$, while the Laplacian eigenvalue equation will be written $\nabla^2\psi=\lambda\psi$, so $\kappa^2=-\lambda$.

By separation of variables into radial and angular coordinates one succeeds in identifying a (complete, orthonormal) set of eigenfunctions $\vert\psi_{nlm}\rangle=\vert\psi_n\rangle\otimes\vert\psi_{lm}\rangle$ within the Hilbert space $L^2\left(\mathbb{S}^3\right)$. This is of course a classical problem in harmonic analysis and mathematical physics, the solution to which is well known in the literature (see, {\it e.g.}\/, refs. \cite{HELGASON, INFELD2, VILENKIN}). The family of radial eigenfunctions $\vert\psi_n\rangle$ will play a distinguished role. The angular eigenfunctions $\vert\psi_{lm}\rangle$ are, of course, the standard spherical harmonics on the unit  2--dimensional sphere $\mathbb{S}^2$. The operator (\ref{landaerre}) is thus represented by the matrix $\langle\psi_{nlm}\vert\Lambda\vert\psi_{n'l'm'}\rangle$. We will see, however, that the relevant physics is captured already at the level of the expectation values $\langle\psi_n\vert\Lambda\vert\psi_{n}\rangle$, without the need to consider the full--blown matrix representing $\Lambda$ in the complete, orthonormal set $\vert\psi_{nlm}\rangle$  within $L^2\left(\mathbb{S}^3\right)$. From the experimentally obtained value of the cosmological constant \cite{PLANCK} one can determine the quantum state 
$\vert\psi_n\rangle$ such that $\langle\psi_n\vert\Lambda\vert\psi_{n}\rangle$ best fits the experimental data. Not surprisingly, the best fit is obtained for a radial state in the semiclassical regime.

This paper is organised as follows. In section \ref{esferico} we summarise the necessary geometrical data pertaining to spherical space $\mathbb{S}^3$. Starting from the metric, the time--independent, free Schroedinger equation is solved by separation of variables, and a complete, orthonormal set of eigenfunctions is identified. Once this set is identified, in section \ref{rraddial} we exploit the power of radial symmetry and compute the matrix representing the cosmological--constant operator. In particular, its expectation value in the radial states $\vert\psi_n\rangle$ is obtained as a function of the radial quantum number $n$. Another result obtained here is the Boltzmann entropy of this Newtonian Universe. In section \ref{piatto} we discuss how to recover the theory on flat space $\mathbb{R}^3$ as a certain limit of our model on $\mathbb{S}^3$. Finally our results are discussed in section \ref{conclusiones}. 

In our use of special functions we follow the conventions of ref. \cite{GRADSHTEYN}.

\section{Spherical space $\mathbb{S}^3$}\label{esferico}

\subsection{Metric, integration measure and Laplacian}

Given the 3--sphere $\mathbb{S}^3$ with radius $R_0$, local coordinates $r,\theta,\varphi$ can be found such that the metric induced by the Euclidean $\mathbb{R}^4$ in which $\mathbb{S}^3$ is embedded reads
\begin{equation}
{\rm d}s^2=\frac{{\rm d}r^2}{1-\frac{r^2}{R_0^2}}+r^2\left({\rm d}\theta^2+\sin^2\theta\,{\rm d}\varphi^2\right),
\label{uno}
\end{equation}
where $0<\theta<\pi$, $0\leq\varphi<2\pi$. Setting $r$ equal to a constant $r_0^2$ we obtain a 2--sphere $\mathbb{S}^2$ with radius $r_0$, while  letting $R_0\to\infty$ one recovers flat space $\mathbb{R}^3$. It is also frequent to use another set of local coordinates $\chi, \theta, \varphi$, where 
\begin{equation}
r=R_0\sin\chi, \qquad 0<\chi<\pi, 
\label{cincuenta}
\end{equation}
and $\theta, \varphi$ remain as above.\footnote{The manifold $\mathbb{S}^3$ being compact, it cannot be covered by a single coordinate chart. Thus, {\it e.g.}\/, the coordinates $\chi, \theta, \varphi$ succeed in covering almost all of $\mathbb{S}^3$, with the exception of a set of measure zero.}   
Then the metric (\ref{uno}) becomes
\begin{equation}
{\rm d}s^2=R_0^2\left[{\rm d}\chi^2+\sin^2\chi\left({\rm d}\theta^2+\sin^2\theta{\rm d}\varphi^2\right)\right].
\label{catorce}
\end{equation}
We identify a radial displacement as corresponding to ${\rm d}\theta=0$, ${\rm d}\varphi=0$. Thus radially symmetric objects will only depend on 
$\chi$. We will also need the Laplacian corresponding to the metric (\ref{catorce}) on $\mathbb{S}^3$:
$$
\nabla^2=\frac{1}{\sqrt{g}}\partial_{i}\left(\sqrt{g}g^{ij}\partial_j\right)
$$
\begin{equation}
=\frac{1}{R_0^2\sin^2\chi}\left[\frac{\partial}{\partial\chi}\left(\sin^2\chi\frac{\partial}{\partial\chi}\right)+\frac{1}{\sin\theta}\frac{\partial}{\partial\theta}\left(\sin\theta\frac{\partial}{\partial\theta}\right)+\frac{1}{\sin^2\theta}\frac{\partial^2}{\partial\varphi^2}\right].
\label{treinta}
\end{equation}
The integration measure ${\rm d}\mu(\chi,\theta,\varphi)$ that is invariant under the isometry group of the metric (\ref{catorce}) is given by
\begin{equation}
{\rm d}\mu(\chi,\theta,\varphi)=\frac{1}{R_0^3}\sqrt{g}\,{\rm d}^3x=\sin^2\chi\sin\theta\,{\rm d}\chi\,{\rm d}\theta\,{\rm d}\varphi.
\label{bolume}
\end{equation}

\subsection{The radial wave equation}

The Laplacian eigenvalue equation $\nabla^2 \psi=\lambda \psi$ on $\mathbb{S}^3$ is found to possess a purely point spectrum given by \cite{HELGASON, INFELD2, VILENKIN} 
\begin{equation}
\lambda_k=-\frac{1}{R_0^2}k(k+2), \qquad k\in\mathbb{N}.
\label{siete}
\end{equation}
The scalar $k(k+2)$ is the eigenvalue of the quadratic Casimir operator of the Lie algebra ${\rm so}(4)$; we observe that  $-\nabla^2$ is a nonnegative operator. The eigenspace corresponding to the eigenvalue $\lambda_k$ has the dimension $(k+1)^2$, as we will see later. In order to obtain the eigenfunctions we separate variables: $\psi(\chi,\theta, \varphi)$ $=$ $R(\chi)Y(\theta,\varphi)$. Then the angular functions are the standard spherical harmonics $Y^{lm}(\theta, \varphi)$ on the unit 2--dimensional sphere $\mathbb{S}^2$, while the radial equation reads
\begin{equation}
\frac{{\rm d}}{{\rm d}\chi}\left(\sin^2\chi\frac{{\rm d}R}{{\rm d}\chi}\right)+\left[\kappa^2R_0^2\sin^2\chi-l(l+1)\right]R=0, \qquad l\in\mathbb{N}.
\label{once}
\end{equation}
Two possibilities present themselves to solve the above radial equation. Each one has its own merits, so we discuss the two of them.

\subsection{Solution in terms of Gegenbauer polynomials}

The change of dependent variable $R(\chi)=\sin^l\chi\,C(\chi)$, followed by the change of independent variable  $x=\cos\chi$, turns the radial equation (\ref{once}) into 
\begin{equation}
(1-x^2)\frac{{\rm d}^2 y}{{\rm d}x^2}-(2l+3)x\,\frac{{\rm d}y}{{\rm d}x}+\left[k(k+2)-l(l+2)\right]\,y(x)=0,
\label{diecisiete}
\end{equation}
where $y(x)=C(\chi)$. The above conforms to the pattern of the Gegenbauer differential equation \cite{GRADSHTEYN},
\begin{equation}
(1-x^2)\frac{{\rm d}^2 y}{{\rm d}x^2}-(2\alpha+1)x\frac{{\rm d}y}{{\rm d}x}+n(n+2\alpha)y(x)=0,
\label{jejenbau}
\end{equation}
with $\alpha=l+1$ and $n=k-l$. 

The solutions to Eq. (\ref{jejenbau}) that are regular at $x=\pm 1$ are polynomials of degree $n$ in $x$ called Gegenbauer polynomials:
\begin{equation}
y(x)=C_n^{\alpha}(x),\qquad n\in\mathbb{N}, \qquad \alpha\geq\frac{1}{2}.
\label{treintados}
\end{equation}
In particular this implies $l\leq k$. The coefficients of the $C_n^{\alpha}(x)$ can be obtained, {\it e.g.}\/, from Eq. (8.932.1) of ref. \cite{GRADSHTEYN}, where these polynomials are expressed in terms of the Gauss hypergeometric function $F(a,b;c;x)$:
\begin{equation}
C_n^{\alpha}(x)=\frac{\Gamma(2\alpha+n)}{\Gamma(n+1)\Gamma(2\alpha)}\,F\left(-n,n+2\alpha;\frac{2\alpha+1}{2};\frac{1-x}{2}\right).
\label{graddy}
\end{equation}
Moreover, Eqs. 7.313.1 and 7.313.2 of ref. \cite{GRADSHTEYN} state the orthogonality property
\begin{equation}
\int_{-1}^1{\rm d}x\,(1-x^2)^{\alpha-1/2}C_n^{\alpha}(x)C_m^{\alpha}(x)=N(n,\alpha)\,\delta_{nm},
\label{treintatres}
\end{equation}
with a normalisation factor that turns out to be
\begin{equation}
N(n,\alpha)=\frac{\pi\,2^{1-2\alpha}\,\Gamma(n+2\alpha)}{n!(\alpha+n)\left[\Gamma(\alpha)\right]^2}.
\label{treintatresbis}
\end{equation}
Therefore the polynomials $G_n^{\alpha}(x)=\left[N(n,\alpha)\right]^{-1/2}\,C_n^{\alpha}(x)$ constitute, for each fixed value of $\alpha\geq 1/2$, a complete, orthonormal set of solutions to Eq. (\ref{jejenbau}) within the Hilbert space $L^2\left([-1,1], {\rm d}\mu_{\alpha}(x)\right)$; the integration measure reads
\begin{equation}
{\rm d}\mu_{\alpha}(x)=(1-x^2)^{\alpha-1/2}{\rm d}x.
\label{quince}
\end{equation}

Since $x=\cos\chi$ and $\alpha=l+1$ we can also write the above measure as
\begin{equation}
{\rm d}\mu_{l+1}(\chi)=-\sin^{2l+2}\chi\,{\rm d}\chi.
\label{vivabiden}
\end{equation}
We will be especially interested in the $l=0$ eigenfunctions, when the above integration measure simplifies to
\begin{equation}
{\rm d}\mu_{1}(\chi)=-\sin^2\chi\,{\rm d}\chi.
\label{vivakamala}
\end{equation}
The negative sign in Eqs. (\ref{vivabiden}), (\ref{vivakamala})  just reflects the fact that the integral (\ref{treintatres}) extends from $x=-1$ to $x=1$, hence $\chi=\arccos x$ runs from $\chi=\pi$ to $\chi=0$. Inverting the integration limits in $\chi$ restores positivity. With the proviso that integration over $\chi$ extend from $\chi=0$ to $\chi=\pi$, we will remove the negative signs from Eqs. (\ref{vivabiden}), (\ref{vivakamala}) and write the integration measure (\ref{bolume}) on the sphere $\mathbb{S}^3$ as
\begin{equation}
{\rm d}\mu(\chi,\theta,\varphi)={\rm d}\mu_1(\chi)\,{\rm d}\mu(\theta,\varphi),\qquad {\rm d}\mu(\theta,\varphi)=\sin\theta\,{\rm d}\theta\,{\rm d}\varphi,
\label{boludo}
\end{equation}
where ${\rm d}\mu(\theta,\varphi)$ is the standard integration measure on the unit 2--sphere $\mathbb{S}^2$. Moreover, the remaining factor $\sin^{2l}\chi$ in the measure (\ref{vivabiden}) must be attached to the eigenfunctions (one factor $\sin^l\chi$ per eigenfunction, as per the change of variables $R(\chi)=\sin^l\chi\,C(\chi)$ applied to the radial equation (\ref{once})).

Altogether, the Laplacian eigenfunctions
\begin{equation}
Y^{klm}(\chi, \theta, \varphi)=2^{l}l!\sqrt{\frac{2(k+1)(k-l)!}{\pi(k+l+1)!}}\,\sin^l\chi\,C_{k-l}^{l+1}(\cos\chi)\,Y^{lm}(\theta,\varphi),
\label{treintacincobis}
\end{equation}
also called hyperspherical harmonics, form an orthonormal
\begin{equation}
\int_{\mathbb{S}^3}{\rm d}\mu_{1}(\chi){\rm d}\mu(\theta, \varphi)\,\left[Y^{klm}(\chi, \theta,\varphi)\right]^*Y^{k'l'm'}(\chi, \theta,\varphi)=\delta^{kk'}\delta^{ll'}\delta^{mm'}
\label{mil}
\end{equation}
and complete
\begin{equation}
\sum_{k=0}^{\infty}\sum_{l=0}^{k}\sum_{m=-l}^l\left[Y^{klm}(\chi, \theta,\varphi)\right]^*Y^{klm}(\chi', \theta',\varphi')=\frac{\delta(\chi-\chi')\delta(\theta-\theta')\delta(\varphi-\varphi')
}{\sin^2\chi\sin\theta}\label{miluno}
\end{equation}
set within the Hilbert space $L^2\left(\mathbb{S}^3,{\rm d}\mu(\chi, \theta, \varphi)\right)$. We see that the range of indices that ensures completeness is $k\in\mathbb{N}$, $l\in\mathbb{N}$ with $l\leq k$, and $m\in\mathbb{Z}$ with $-l\leq m\leq l$. 

Now the 3--dimensional sphere $\mathbb{S}^3$ is the homogeneous manifold ${\rm SO}(4)/{\rm SO}(3)$ \cite{HELGASON,  VILENKIN}, and the collection of all hyperspherical harmonics $Y^{klm}$ provides the basis vectors for a unitary representation of the Lie algebra ${\rm so}(4)$. The carrier space for this representation is $L^2\left(\mathbb{S}^3,{\rm d}\mu(\chi, \theta, \varphi)\right)$. This representation is reducible, its irreducible components being the subspaces spanned by those states $Y^{klm}$ with fixed $k$, with $l\leq k$, and with $-l\leq m\leq l$. One readily verifies that these latter subspaces are $(k+1)^2$--dimensional.

\subsection{Solution in terms of ladder operators}

Next we present an alternative resolution of the radial equation (\ref{once}), one that does not have recourse to the Gegenbauer differential equation and the corresponding polynomials. Instead it uses the method of ladder operators developed in ref. \cite{INFELD2}.

By inspection, the function
\begin{equation}
\bar R_{p}^1(\chi)=\frac{\sin(p+1)\chi}{\sin\chi}, \qquad p\in\mathbb{R} 
\label{mildos} 
\end{equation}
satisfies the radial equation (\ref{once}) with $l=0$ whenever $\kappa$ and $p$ are such that
\begin{equation}
\kappa^2R_0^2=p(p+2).
\label{miltres}
\end{equation}
Since $\kappa^2=-\lambda$, comparison of (\ref{miltres}) with the allowed eigenvalues (\ref{siete}) enforces $p=k\in\mathbb{N}$, and the family of solutions (\ref{mildos}) becomes
\begin{equation}
\bar R_{k}^1(\chi)=\frac{\sin(k+1)\chi}{\sin\chi}, \qquad k\in\mathbb{N}. 
\label{milcuatro} 
\end{equation}
We claim that the function $\bar R_{k}^{l+1}(\chi)$ defined as
\begin{equation}
\bar R_{k}^{l+1}(\chi)=\sin^l\chi\,\left(\frac{1}{\sin\chi}\frac{{\rm d}}{{\rm d}\chi}\right)^l\bar R_{k}^{1}(\chi), \qquad l\leq k 
\label{milcinco}
\end{equation}
satisfies the radial equation (\ref{once}) with an arbitrary value of $l\in\mathbb{N}$.

In order to prove the above statement, we return to the radial equation (\ref{once}) and perform the first change of variables mentioned there: $\bar R_{k}^{l+1}(\chi)=\sin^l\chi\,\bar C_{k}^{l+1}(\chi)$. Then the radial equation becomes
\begin{equation}
\frac{{\rm d}^2\bar C_{k}^{l+1}}{{\rm d}\chi^2}+2(l+1)\cot\chi\,\frac{{\rm d}\bar C_{k}^{l+1}}{{\rm d}\chi}+\left[k(k+2)-l(l+2)\right]\,\bar C_{k}^{l+1}(\chi)=0.
\label{diecisietebis}
\end{equation}
Differentiating the above once more yields
\begin{equation}
\frac{{\rm d}^3\bar C_{k}^{l+1}}{{\rm d}\chi^3}+2(l+1)\cot\chi\frac{{\rm d}^2\bar C_{k}^{l+1}}{{\rm d}\chi^2}+\left[k(k+2)-l(l+2)-2(l+1)\csc^2\chi\right]\frac{{\rm d}\bar C_{k}^{l+1}}{{\rm d}\chi}=0,
\label{milseis}
\end{equation}
and a little algebra verifies that the ansatz
\begin{equation}
\frac{{\rm d}\bar C_{k}^{l+1}}{{\rm d}\chi}=\sin\chi\,\bar C_{k}^{l+2},
\label{ansas}
\end{equation}
when substituted into Eq. (\ref{milseis}), reproduces the radial equation (\ref{diecisietebis}), but with $l+1$ replacing $l$. Therefore the ansatz (\ref{ansas}) is correct, and by induction we can write
\begin{equation}
\bar C_{k}^{l+1}(\chi)=\left(\frac{1}{\sin\chi}\frac{{\rm d}}{{\rm d}\chi}\right)^l\bar C_{k}^1(\chi), \qquad l\leq k.
\label{milsiete}
\end{equation}
In this way the eigenfunction $\bar C_{k}^1(\chi)$ plays the role of a vacuum state for the angular momentum $l$, and the operator $(1/\sin\chi)({\rm d}/{\rm d}\chi)$ acts as a ladder operator. Modulo normalisation, the complete radial eigenfunction reads
\begin{equation}
\bar R_{k}^{l+1}(\chi)=\sin^l\chi\,\bar C_{k}^{l+1}(\chi)=\sin^l\chi\,\left(\frac{1}{\sin\chi}\frac{{\rm d}}{{\rm d}\chi}\right)^l\bar C_{k}^1(\chi), \qquad l\leq k.
\label{lockhimup}
\end{equation}
as claimed above.

The bars on top of $\bar R_{k}^{l+1}$, $\bar C_{k}^{l+1}$ are meant to remind us that we are disregarding normalisations. Even if the vacuum state $\bar C_{k}^1(\chi)$ were normalised to unity, its excitations $\bar C_{k}^{l+1}(\chi)$ need not (and generally will not) be normalised to unity, so correct normalisation must be checked after each application of the ladder operator $(1/\sin\chi)({\rm d}/{\rm d}\chi)$. Since we already know that $\bar C_{k}^{l+1}(\chi)$ must be proportional to the Gegenbauer polynomial $C_{k}^{l+1}(\chi)$ anyway, it is more practical to resort to Eq. (\ref{treintatresbis}) for normalisation.

\section{Radial symmetry}\label{rraddial}

\subsection{Operators}

Three relevant quantum operators that are radially symmetric are the Newtonian potential $V$, the cosmological constant $\Lambda$, and the Boltzmann entropy $S$.

We begin with the Newtonian potential $V$. A point mass located at the origin generates a Newtonian potential that is purely radial, so it satifies the radial part of the Laplace equation. For this we set $\kappa^2=0=l$ in Eq. (\ref{once}) and get
\begin{equation}
\frac{{\rm d}}{{\rm d}\chi}\left(\sin^2\chi\frac{{\rm d}V}{{\rm d}\chi}\right)=0,
\label{trentino}
\end{equation}
after renaming $R$ as $V$. The general solution to (\ref{trentino}) reads
\begin{equation}
V(\chi)=c_1\cot(\chi)+c_2 \qquad c_1,c_2\in\mathbb{R}.
\label{altoadige}
\end{equation}
Hence the function (\ref{altoadige}) is the Newtonian potential on $\mathbb{S}^3$ \cite{INFELD2}. Upon setting $c_2=0$, $c_1=1/R_0$ and  expanding around $\chi=0$ we obtain $V(\chi)=1/(R_0\chi)$, that is $V(r)=1/r$, which is of course the Newtonian potential in flat space $\mathbb{R}^3$.

The cosmological constant is represented quantum--mechanically as an operator that, in flat space, is proportional to the inverse of the squared radial distance $\rho$ \cite{NOSALTRES}. In $\mathbb{R}^3$, the standard choice is of course the centrifugal potential energy,
\begin{equation}
U_{\mathbb{R}^3}(\rho)=\frac{{\bf L}^2}{2M\rho^2}.
\label{zentri}
\end{equation}
The analogous centrifugal potential energy in $\mathbb{S}^3$ reads \cite{INFELD2}
\begin{equation}
U_{\mathbb{S}^3}(r)=\frac{{\bf L}^2}{2MR_0^2\sin^2\chi},
\label{fuga}
\end{equation}
which suggests considering the operator
\begin{equation}
\Lambda=\frac{C}{R_0^2\sin^2\chi}
\label{cincuentacinco}
\end{equation}
as the quantum--mechanical operator that will represent the cosmological constant. The numerical constant $C$, a dimensionless proportionality factor, will be determined presently. We will not set $C$ proportional to the square of the angular momentum because we would like radially symmetric states, for which ${\bf L}^2=0$, to contribute to the cosmological constant.

{}Finally we turn our attention to the entropy operator. In the entropic approach to gravity \cite{VERLINDE}, gravitational equipotential surfaces have been shown to be isoentropic surfaces. Using this property, in refs. \cite{CABRERA, NOSALTRESTRIS} we have computed the Boltzmann entropy of a flat, Newtonian Universe $\mathbb{R}\times\mathbb{R}^3$ when the matter contents of the Universe is represented by a quantum state $\vert\psi\rangle$; the entropy operator considered reads $S={\cal N}k_BMH_0\rho^2/\hbar$. Here $k_B$ is Boltzmann's constant, $M$ is the total mass (baryonic and dark) of the observable Universe, $H_0$ is Hubble's constant, and $\rho$ is a radial coordinate with the dimensions of length. A dimensionless numerical factor ${\cal N}$ is left undetermined by our argument. However, on general grounds, we expect ${\cal N}$ to be of order unity. The role of $\rho$ in flat space is played on the sphere $\mathbb{S}^3$ by $r$, as per Eq. (\ref{cincuenta}). Thus the entropy operator on $\mathbb{S}^3$ must be
\begin{equation}
S={\cal N}\,\frac{k_BMH_0}{\hbar}R_0^2\sin^2\chi.
\label{cuatrocientosdos}
\end{equation}

\subsection{Wavefunctions}

We will call {\it radially symmetric eigenfunctions}\/ those $Y^{klm}$ with $m=0=l$. Since $Y^{00}(\theta, \varphi)=(4\pi)^{-1/2}$, the eigenfunction $Y^{k00}$ depends only on $\chi$. Now $n=k-l$ and $l=0$, so the two quantum numbers $n$ and $k$ are equal when dealing with radially symmetric wavefunctions. Applying Eq. 8.937.1 of ref. \cite{GRADSHTEYN}, reproduced below for convenience,
\begin{equation}
C_k^1(\cos\chi)=\frac{\sin(k+1)\chi}{\sin\chi},
\label{comenencia}
\end{equation}
we arrive at
\begin{equation}
Y^{k00}(\chi)=\frac{\sqrt{2}}{2\pi}C_k^1(\cos\chi)=\frac{\sqrt{2}}{2\pi}\frac{\sin(k+1)\chi}{\sin\chi},  \qquad k\in\mathbb{N}.
\label{graddybis}
\end{equation}
On the right--hand side above we recognise the Weyl character ${\rm ch}_{k+1}(\chi)$ of the $(k+1)$--dimensional irreducible representation of the Lie algebra ${\rm su}(2)$. Hence
\begin{equation}
Y^{k00}(\chi)=\frac{\sqrt{2}}{2\pi}\,{\rm ch}_{k+1}(\chi), \qquad k\in\mathbb{N}.
\label{doscientosuno}
\end{equation}
This should come as no surprise, because $\mathbb{S}^3$ is diffeomorphic as a manifold to the Lie group ${\rm SU}(2)$, and because Weyl characters are eigenfunctions of the Laplacian on the Lie algebra \cite{HELGASON, VILENKIN}.

\subsection{The cosmological constant}

The expectation value of the operator (\ref{cincuentacinco}) in the radially symmetric states (\ref{doscientosuno}) can be readily computed:
\begin{equation}
\langle Y^{k00} \vert\Lambda\vert Y^{k00}\rangle=\frac{2C}{R_0^2}(k+1).
\label{milonce}
\end{equation}
The dimensionless numerical constant $C$, so far undetermined, can now be picked positive to ensure a positive cosmological constant for spherical space. This is in agreement with current experimental data \cite{PLANCK} and also with the important fact that, in Newtonian terms, one expects $C<0$ for an attractive force but $C>0$ for a repulsive force (such as the cosmological constant). For definiteness we will make the specific choices
\begin{equation}
C=1, \qquad R_0=R_U=4.4\times 10^{26} {\rm m}, 
\label{choix}
\end{equation}
where  $R_U$ is the radius of the observable Universe.  We obtain the table
\begin{equation}
\begin{tabular}{| c | c | c|}\hline
$\langle Y^{k00}\vert\Lambda\vert Y^{k00}\rangle$ & in m{}$^{-2}$  &  in Planck units  \\ \hline
$k=0$ & $1.0\times 10^{-53}$ &  $2.5\times 10^{-123}$  \\ \hline
$k=1$ & $2.0\times 10^{-53}$ & $5.1\times 10^{-123}$ \\ \hline
$\vdots$ & $\vdots$ & $\vdots$   \\ \hline
$k=9$ & $1.0\times 10^{-52}$ & $2.5\times 10^{-122}$ \\ \hline
$k=10$ & $1.1\times 10^{-52}$ & $2.8\times 10^{-122}$ \\ \hline
\end{tabular}
\label{trumpthebullshitter}
\end{equation}
We see that the best fit to the current data $\Lambda=1.1\times 10^{-52}$ (in ${\rm m}^{-2}$) or $\Lambda=2.8\times 10^{-122}$ (in Planckian units) \cite{PLANCK} is attained for a value of the radial quantum number $k=10$, well into the semiclassical regime as expected.

We can further compute the matrix representing the cosmological constant within the subspace of radially symmetric states. It turns out to be
\begin{equation}
\langle Y^{k00} \vert\Lambda\vert Y^{k'00}\rangle=\frac{2}{R_U^2}\,{\rm min}(k+1,k'+1)\,\delta(P_k,P_{k'}), \qquad k,k'\in\mathbb{N},
\label{jailtrump}
\end{equation}
where $\delta(P_k,P_{k'})$ equals 1 (resp. zero) if $k$ and $k'$ have the same (resp. opposite) parity.

To round up our analysis it is possible to obtain the full matrix $\langle Y^{klm}\vert\Lambda\vert Y^{k'l'm'}\rangle$ without restriction to radially symmetric states. Unfortunately the result, although proportional to $\delta^{ll'}\delta^{mm'}$, is too involved and too unenlightening to be of any practical use, so we omit quoting it.

\subsection{Boltzmann entropy}

Here again one can compute the matrix $\langle Y^{klm}\vert S\vert Y^{k'l'm'}\rangle$ representing the entropy operator in the full Hilbert space $L^2\left(\mathbb{S}^3, {\rm d}\mu(\chi,\theta,\varphi)\right)$. Again this matrix will be proportional to $\delta^{ll'}\delta^{mm'}$. However, for the same reasons as before, we will restrict our attention to radially symmetric states. Moreover, the expectation value $\langle Y^{k00}\vert S\vert Y^{k00}\rangle$ will suffice to obtain an estimate for an order of magnitude. Thus using the eigenfunctions (\ref{graddybis}) and the operator (\ref{cuatrocientosdos}) one arrives at the expectation value 
\begin{equation}
\langle Y^{k00}\vert S\vert Y^{k00}\rangle={\cal N}\frac{k_BMH_0R_U^2}{2\hbar},
\label{cuatrocientostres}
\end{equation}
which turns out to be independent of the radial quantum number. Taking ${\cal N}$ of order unity and substituting the current  values of the cosmological data \cite{PLANCK} we obtain 
\begin{equation}
\langle Y^{k00}\vert S\vert Y^{k00}\rangle\simeq 10^{123}k_B;
\label{cuatrocientoscuatro}
\end{equation}
a result which certainly saturates, but does not violate, the upper bound $10^{123}k_B$ set by the holographic principle \cite{CABRERA, NOSALTRESTRIS}.

\section{The flat--space limit}\label{piatto}

In this section we verify that the results obtained in $\mathbb{S}^3$ correctly reduce to their counterparts in $\mathbb{R}^3$ \cite{CABRERA, NOSALTRES, NOSALTRESTRIS} as one lets $R_0\to\infty$.

After the change of variables (\ref{cincuenta}), the limit $R_0\to\infty$ is to be understood as $\chi\to 0$, hence we can systematically replace $\sin\chi$ with $\chi$. This is a rule of thumb that should be judiciously applied. For example, the numerator $\sin(k+1)\chi$ in the eigenfunction (\ref{graddybis}) may {\it not}\/ be approximated by $(k+1)\chi$ (even less so in the semiclassical limit $k\to\infty$).

{}Following the above recipe we may safely state the following: as $\chi\to 0$,\\ 
{\it i)} the integration measure (\ref{bolume}) on $\mathbb{S}^3$ becomes the usual integration measure on
$\mathbb{R}^3$;\\
{\it ii)} the radial equation (\ref{once}) on $\mathbb{S}^3$ correctly reduces to its counterpart in $\mathbb{R}^3$;\\
{\it iii)} the radially symmetric eigenfunctions (\ref{graddybis}) correctly reduce to their flat--space counterparts. This is best seen at the level of Eq. (\ref{milcuatro}), where
\begin{equation}
\bar R_{k}^1(\chi)=\frac{\sin(k+1)\chi}{\sin\chi}\to\frac{\sin(k+1)\chi}{\chi}.
\label{mildiez} 
\end{equation}
The above are the (unnormalised) radially symmetric eigenfunctions used in the flat--space analysis of ref. \cite{NOSALTRES}. Quantisation of the radial momentum $k$ in integer multiples of $\pi/R_U$ arises in flat space upon introducing a boundary condition at $R_U$. Then the allowed energy levels in $\mathbb{S}^3$ become those expected for $\mathbb{R}^3$ if one maintains quadratic terms in $k$ while dropping linear terms;\\
{\it iv)} unnormalised, nonradially symmetric, radial eigenfunctions $\bar R_{k}^l(\chi)$ (with $l>0$) have been obtained in Eq. (\ref{milcinco}) by the action of ladder operators. Now as $\chi\to 0$ the latter become
\begin{equation}
\frac{1}{\sin\chi}\frac{{\rm d}}{{\rm d}\chi}\to\frac{1}{\chi}\frac{{\rm d}}{{\rm d}\chi},
\label{cuatrocientossiete}
\end{equation}
and one recovers the flat--space ladder operators needed to obtain radial eigenfunctions with $l>0$ from the radially symmetric ``vacuum" $\bar R_{k}^1(\chi)$;\\
{\it v)} last but not least, the 2--dimensional spherical harmonics  $Y^{lm}(\theta, \varphi)$ remain the same.

\section{Discussion}\label{conclusiones}

In this paper we have studied the Newtonian limit of a spacetime given by the product of the time axis $\mathbb{R}$, times the 3--dimensional sphere $\mathbb{S}^3$, the latter endowed with the usual round metric. This is the spacetime manifold considered by Einstein in his attempt to construct a static cosmological model \cite{TOLMAN}. Despite the many shortcomings of this cosmology, it exhibits some interesting features that merit analysis in  this simplified setup, if only with the aim of applying them to more realistic approaches later. 

As explained in ref. \cite{NOSALTRES}, the weak--gravity limit of the cosmological fluid in $\mathbb{R}\times\mathbb{M}^3$, where 
$\mathbb{M}^3$ is a spatial 3--manifold, can be mimicked by the probability fluid of the nonrelativistic quantum mechanics of a single particle obeying the Schroedinger equation in $\mathbb{R}\times\mathbb{M}^3$. For simplicity we have considered the case of a free particle. Hence energy eigenfunctions coincide with Laplacian eigenfunctions on the manifold $\mathbb{M}^3$, in our case $\mathbb{S}^3$. 

To begin with, we haved summarised the derivation of the Laplacian eigenfunctions on $\mathbb{S}^3$ known in the literature as hyperspherical harmonics \cite{HELGASON, INFELD2, VILENKIN}. This done, our major interest lay in obtaining an estimate for the cosmological constant. This estimate has been obtained as the expectation value of a suitably defined operator, taken in a suitably selected quantum eigenfunction. The latter would then represent the quantum state that our hypothetical Newtonian Universe $\mathbb{R}\times\mathbb{S}^3$ finds itself in. It turns out that the best fit to the experimentally measured value for $\Lambda$ is attained for a radial eigenfunction lying deep inside the semiclassical regime. 

The notion of a cosmological constant that scales like the inverse of the squared radius of the Universe, as our operator (\ref{cincuentacinco}), dates back to Einstein's static Universe. In Newtonian terms, the cosmological constant can be understood as being analogous to a centrifugal force that, however, is nonvanishing even when the angular momentum ${\bf L}$ vanishes. Of course, the centrifugal force vanishes identically whenever ${\bf L}={\bf 0}$; the analogy breaks down at this point. This notwithstanding, a central potential that in flat space scales like $r^{-2}$ mimics the classical centrifugal potential ${\bf L}^2/2Mr^2$. From this standpoint, our choice (\ref{cincuentacinco})  for the cosmological constant $\Lambda$ in spherical space is physically meaningful. 

We also observe that, up to multiplicative physical constants, the operator representing (\ref{cincuentacinco}) the cosmological constant and the operator (\ref{cuatrocientosdos}) representing the Boltzmann entropy are mutually inverse. This property has already been noticed to hold in flat space \cite{NOSALTRES}; it reflects the fact that the experimentally measured cosmological constant $10^{-122}$ (in Planck units) lies quite close to the inverse of the upper bound $10^{123}$ set by the holographic principle for the (dimensionless) entropy $S/k_B$. 

At first sight, our view of the Newtonian cosmological constant as represented by the operator (\ref{cincuentacinco}) might appear to deviate from standard lore, where $\Lambda$ leads to a modified Poisson equation through the addition of a negative mass density. Instead, our Newtonian potential $V$ continues to satisfy the usual Poisson equation (\ref{trentino}), while an additional ``centrifugal force" represented by the operator (\ref{cincuentacinco}) contributes to the accelerated expansion of the Universe. Upon closer inspection, however, our viewpoint ($\Lambda$ as a centrifugal force, hence repulsive) turns out to be equivalent to the standard viewpoint ($\Lambda$ as a modification of the Poisson equation by a negative mass density).

The spatial manifold $\mathbb{S}^3$ is diffeomorphic to the Lie group ${\rm SU}(2)$. Correspondingly, the hyperspherical harmonics (\ref{treintacincobis}) depend on the three coordinates $\chi,\theta, \varphi$ parametrising ${\rm SU}(2)$. We have found that radial symmetry provides a substantial technical simplification, while at the same time providing us with physically meaningful estimates for the cosmological constant and the Boltzmann entropy that would be much harder to obtain in the absence of radial symmetry. The reason this simplification works is that the operators (\ref{cincuentacinco}) and (\ref{cuatrocientosdos}) representing the cosmological constant and the Boltzmann entropy are themselves radially symmetric. From a Lie--group theoretic point of view, the restriction to the radially symmetric wavefunctions (\ref{doscientosuno}) corresponds to considering the {\it effective Abelian theory}\/ described by the maximal torus ${\rm U}(1)$, instead of the full--blown nonabelian group ${\rm SU}(2)$.

Another interesting feature of our model is the following. While the cosmological--constant operator on flat space needed regularisation  \cite{NOSALTRES}, {\it positive curvature does away with the need to regularise}\/. The Hadamard regularisation \cite{BLANCHET} we used in flat space $\mathbb{R}^3$ to properly define the expectation value of the cosmological--constant operator becomes unnecessary in the sphere $\mathbb{S}^3$. Now flat space $\mathbb{R}^3$ qualifies as a Lie group as much as $\mathbb{S}^3$, but it is Abelian. We see that the simplification provided by flatness and the Abelian property is offset by the need to regularise. Since all simple Lie groups qualify as Einstein manifolds \cite{GOLDBERG}, we can expect a similar behaviour when dealing with higher--rank groups such as, {\it e.g.}\/, ${\rm SU}(n)$. It is intriguing to muse on an eventual sum over dimensions \cite{CFI} in such spacetimes, where nonabelianity plays a role.

Despite arduous efforts over many decades, the scientific community does not yet have a microscopic theory to explain the physics of the cosmological constant and the fabric of spacetime. One can nonetheless come a long way  without making explicit assumptions about the microscopic degrees of freedom of spacetime and about the nature of dark energy; indeed such a thermodynamic approach can already claim relevant successes \cite{PADDY2}. The Newtonian model presented here can be regarded as a contribution to this thermodynamic approach.

\vskip0.5cm
\noindent
{\bf Acknowledgements} This research was supported by grant no. RTI2018-102256-B-I00 (Spain).

\end{document}